\documentclass{sig-alternate-05-2015}

\begin{document}

% Copyright
\setcopyright{acmcopyright}
%\setcopyright{acmlicensed}
%\setcopyright{rightsretained}
%\setcopyright{usgov}
%\setcopyright{usgovmixed}
%\setcopyright{cagov}
%\setcopyright{cagovmixed}

% DOI
\doi{10.475/123_4}

% ISBN
\isbn{123-4567-24-567/08/06}

%Conference
\conferenceinfo{XSEDE '16}{July 17--23, 2016, Miami, FL, USA}

\acmPrice{\$15.00}

%
% --- Author Metadata here ---
\conferenceinfo{XSEDE}{'16 Miami, FL USA}
%\CopyrightYear{2007} % Allows default copyright year (20XX) to be over-ridden - IF NEED BE.
%\crdata{0-12345-67-8/90/01}  % Allows default copyright data (0-89791-88-6/97/05) to be over-ridden - IF NEED BE.
% --- End of Author Metadata ---

\title{Building a Shared Resource HPC Center Across University Schools and Institutes: A Case Study}
%%\subtitle{[Extended Abstract]
%%\titlenote{A full version of this paper is available as
%%\textit{Author's Guide to Preparing ACM SIG Proceedings Using
%%\LaTeX$2_\epsilon$\ and BibTeX} at
%%\texttt{www.acm.org/eaddress.htm}}}

\numberofauthors{11} %  in this sample file, there are a *total*
\author{
% 1st. author
\alignauthor
Glen MacLachlan\\
       \affaddr{Office of Technology Services}\\
       \affaddr{Columbian College of Arts and Sciences}\\
       \affaddr{The George Washington University}\\
       \affaddr{725 21st Street, NW}\\
       \affaddr{Washington, DC 20052}\\
       \email{maclach@gwu.edu}
% 2nd. author
\alignauthor
Jason Hurlburt\\
       \affaddr{SEAS Computing Facility}\\
       \affaddr{School of Engineering and Applied Sciences}\\
       \affaddr{The George Washington University}\\
       \affaddr{800 22nd Street, NW}\\
       \affaddr{Washington, DC 20052}\\
       \email{hurlburj@gwu.edu}
% 3rd. author
\alignauthor Marco Suarez\\
       \affaddr{SEAS Computing Facility}\\
       \affaddr{School of Engineering and Applied Sciences}\\
       \affaddr{The George Washington University}\\
       \affaddr{800 22nd Street, NW}\\
       \affaddr{Washington, DC 20052}\\
       \email{suarezm@gwu.edu}
\and  % use '\and' if you need 'another row' of author names
% 4th. author
\alignauthor Kai Leung Wong\\
       \affaddr{Division of Information Technology}\\
       \affaddr{The George Washington University}\\
       \affaddr{45085 University Drive}\\
       \email{aklwong@gwu.edu}
       % 5th. author
\alignauthor William Burke\\
       \affaddr{Division of Information Technology}\\
       \affaddr{The George Washington University}\\
       \affaddr{44983 Knoll Square}\\
       \affaddr{Ashburn, VA 20147}\\
       \email{wburke@gwu.edu}
% 6th. author
\alignauthor Terrence Lewis\\
       \affaddr{Division of Information Technology}\\
       \affaddr{The George Washington University}\\
       \affaddr{44983 Knoll Square}\\
       \affaddr{Ashburn, VA 20147}\\
       \email{tlewis@gwu.edu}
\and  % use '\and' if you need 'another row' of author names
% 5th. author
\alignauthor Andrew Gallo\\
       \affaddr{Division of Information Technology}\\
       \affaddr{The George Washington University}\\
       \affaddr{44983 Knoll Square}\\
       \affaddr{Ashburn, VA 20147}\\
       \email{agallo@gwu.edu}
% 6th. author
\alignauthor Jaroslav Flidr\\
       \affaddr{Division of Information Technology}\\
       \affaddr{The George Washington University}\\
       \affaddr{44983 Knoll Square}\\
       \affaddr{Ashburn, VA 20147}\\
       \email{jflidr@gwu.edu}
 % use '\and' if you need 'another row' of author names
% 8th. author
\alignauthor Raoul Gabiam\\
       \affaddr{SEAS Computing Facility}\\
       \affaddr{School of Engineering and Applied Sciences}\\
       \affaddr{The George Washington University}\\
       \affaddr{800 22nd Street, NW}\\
       \affaddr{Washington, DC 20052}\\
       \email{gabiamr@gwu.edu}
% 9th. author
\and
\alignauthor Janis Nicholas\\
       \affaddr{Office of Technology Services}\\
       \affaddr{Columbian College of Arts and Sciences}\\
       \affaddr{The George Washington University}\\
       \affaddr{801 22nd Street, NW}\\
       \affaddr{Washington, DC 20052}\\
       \email{jross16@gwu.edu}
% 10th. author
\alignauthor Brian Ensor\\
       \affaddr{Division of Information Technology}\\
       \affaddr{The George Washington University}\\
       \affaddr{44983 Knoll Square}\\
       \affaddr{Ashburn, VA 20147}\\
       \email{bensor@gwu.edu}
}
% There's nothing stopping you putting the seventh, eighth, etc.
% author on the opening page (as the 'third row') but we ask,
% for aesthetic reasons that you place these 'additional authors'
% in the \additional authors block, viz.
%%\additionalauthors{Additional authors: John Smith (The Th{\o}rv{\"a}ld Group,
%%email: {\texttt{jsmith@affiliation.org}}) and Julius P.~Kumquat
%%(The Kumquat Consortium, email: {\texttt{jpkumquat@consortium.net}}).}
\date{29 April 2016}
% Just remember to make sure that the TOTAL number of authors
% is the number that will appear on the first page PLUS the
% number that will appear in the \additionalauthors section.

\maketitle
\begin{abstract}
Over the past several years, The George Washington University~\cite{gwu} has recruited a significant number of researchers in a wide variety of domains requiring the availability of advanced computational resources. We discuss the challenges and obstacles encountered planning and establishing a first-time high performance computing center at the university level and present a set of solutions that will be useful for any university developing a fledgling high performance computing center. We focus on justification and cost model, strategies for determining anticipated use cases, planning appropriate resources, staffing, user engagement, and metrics for gauging success. 
\end{abstract}

\begin{CCSXML}
<ccs2012>
<concept>
<concept_id>10003456.10003457.10003490</concept_id>
<concept_desc>Social and professional topics~Management of computing and information systems</concept_desc>
<concept_significance>500</concept_significance>
</concept>
</ccs2012>
\end{CCSXML}

\ccsdesc[500]{Social and professional topics~Management of computing and information systems}


%
%  Use this command to print the description
%
\printccsdesc

% We no longer use \terms command

\keywords{information systems; management}

\section{Introduction}
Over the past several years, The George Washington University (GWU) has recruited a significant number of faculty in a wide variety of domains such as Biology, 
Engineering, Genomics, Physics, 
Chemistry, and Statistics whose research efforts depend heavily upon the availability of advanced computational resources. 
There are over 25,000 enrolled students and more than 100 
different departments and programs at the University. GWU is similar to other institutions that aim to increase 
their research profiles in that it must simultaneously increase its investment in faculty and high performance computing (HPC) infrastructure. 
Until the availabity of a local high performance computing center (HPCC), our faculty have used external HPC resources and assembled small, standalone clusters to serve the specific 
needs of their research groups. The procurement, installation, management, and operational support of these clusters are typically the responsibility of the individual research groups who 
own them. Operationally, this leaves the full burden of maintaining clusters and securing data on faculty and graduate students. Given the many commonalities among these individual clusters, 
including computational functionality and operational problems, the prospect of centralizing HPC resources into a common, shared cluster infrastructure presents a substantial economy of scale. This economy of scale benefits both the College and individual faculty member. The University benefits from operating cost efficiencies, while faculty and research groups benefit by having
access to larger computational resources with burst capabilities beyond that which an individual cluster would provide, with the additional benefit of having the HPC resources 
managed professionally by dedicated IT staff.

This manuscript is the corpus of knowledge acquired in planning, provisioning, and maintaining a jointly owned and shared HPCC across schools and institutes in a university environment. 
Lessons learned include justifying the initial and ongoing investment in HPC resources and staffing. We focus on several key areas: 
\begin{enumerate}
\item Key Design Decisions
\item Initial Project Budget
\item Investment and Expansion
\item Schedule and Milestones
\item User Support and Engagement
\end{enumerate}

\section{Colonial One Background} 
Colonial One~\cite{colonialone} is the flagship HPC cluster at GWU and is owned by various stakeholders within the University. See Table~\ref{tab:userdisp} for the current
breakdown of stakeholders. Colonial One is a Dell cluster with the following current specifications:
\begin{itemize}
\item 213 Dell C8220 compute nodes
\begin{itemize}
\item 159 CPU nodes with dual Intel 8-core E5-2670 CPUs
\item 53 GPU nodes with dual NVIDIA Kepler K20 GPUs and dual Intel 6-core E5-2620 CPUs
\item 1 large memory 2TB node with quad 12-Core 3.0GHz Xeon E7-8857v2 CPUs
\end{itemize}
\item Mellanox FDR Infiniband
\item 1/4 PB NFS primary storage 
\item 1/4 PB Intel Enterprise Edition Lustre fast scratch storage 
\item 1/4 PB Dell Compellent long-term archival storage
\end{itemize}

\begin{table}
\centering
\caption{Stakeholder Shares by Discipline}
\begin{tabular}{|l|c|} \hline
Stakeholder& Portion Owned  (\%)\\ \hline
Arts \& Sciences & 52\\ \hline
Engineering \& Applied Sciences & 19\\ \hline
Computional Biology& 17\\ \hline
Public Health& 3\\ \hline
Open Shares& 9\\ \hline
\hline\end{tabular}
\label{tab:userdisp}
\end{table}

Additionally, Colonial One has three login nodes for redundancy and load-balancing and four additional compute nodes reserved for support staff for testing and debugging.   
 
The user-base served by Colonial One is diverse, representing more than 430 researchers in more than 110 research groups including physical sciences, life sciences, engineering, 
and economics, with \$31 million in grants awarded. The cluster processes more than 2,000 jobs per day and operates continuously throughout the year. Access to the cluster open to the University community at no charge 
and priority in a fair-share model is possible by investment in the cluster with priority being proportional to the investment made. We forego the condo model of ownership in favor of 
fair-share primarily because extending burst capacity to researchers is natural in a fair-share environment but unnecessarily complicated in the condo model.  

An HPC Advisory Committee is responsible for Long-term planning and a faculty representative has been appointed from each School and Institute and is chaired by the Director of Research
Technology Services. The Committee advocates for faculty interests and makes recommendations to The Office of the Vice President for Research (OVPR) and the Board of Trustees. 

\section{Key Design Decisions}
There are multiple approaches to building high performance computing clusters. Small clusters can be built using simple desktop computers operating in common office and lab areas, 
while larger clusters are built using enterprise-grade rack-mountable servers hosted in robust data center environments. Our goal was to design a shared cluster that is heterogeneous in 
terms of hardware configuration (varying node configurations supporting varying use cases) though homogenous in terms of hardware platform.

\subsection{Interconnect Network Fabric}
Because of the diversity of use cases, the cluster is implemented using an FDR InfiniBand (IB) network. While first-time cluster users and small-scale jobs would 
run efficiently on a simple 1 Gbps Ethernet network, MPI users require extremely fast interconnects for several types of computation. Because of this requirement, the cluster has been 
designed to meet the high-end needs of researchers. The FDR IB network fabric provides 56 Gbps of low-latency throughput to each node, versus 1 Gbps or 10 Gbps over Ethernet network fabrics. 
As InfiniBand is a specialized network technology, there are very few manufacturers of IB switch hardware. Presently, Mellanox is the only manufacturer that offers FDR IB switches. 
As such, there is no competitive analysis by manufacturer, but multiple bids were solicited from various Mellanox resellers.

GWU considered the full cost three network interconnect options, and compared the total cost per client (client network interface card, switch port, and patch cable). Gigabit Ethernet 
is extremely inexpensive, while 10 gigabit Ethernet and FDR InfiniBand are similar in cost. While FDR InfiniBand is the highest cost technology, its performance of 56 Gbps is more than 
five times faster than 10 gigabit Ethernet though there is a greater cost per client. While it is technically feasible to design clusters with multiple interconnect network fabrics 
(i.e. some nodes using standard 1 Gbps Ethernet, and others utilizing InfiniBand), this introduces complexity in the system design as well as operational support.

Based on the total number of ports required, a fat tree network topology with smaller distributed switches is utilized, as opposed to a more centralized chassis-based topology. 
The IB fabric's fat tree topology with core and edge/top of rack switches was selected based on entry cost and modularity in deployment compared to that of chassis-based systems. 
Given the cluster's size and use cases, a large chassis-based architecture would add significantly to the initial investment cost, and provide little return in terms of performance. 
The proposed topology provides for full, non-blocking throughput for the storage systems, and 2:1 oversubscription for the compute nodes. Oversubscription in HPC IB fabrics is not 
uncommon, and refers to the ratio of allocated bandwidth to guaranteed bandwidth per compute node. Implementing the IB fabric as fully non-blocking for both storage systems and compute 
nodes would require a higher quantity of IB switches.

\subsection{Storage}
There are generally three tiers of storage associated with HPC clusters; primary, scratch, and archive. All three tiers are important for end users throughout the process of 
running HPC jobs.
\subsubsection{Primary Storage}
The primary storage system for the cluster is used for end user data in a staging phase prior to running analysis, or for data results at rest after the completion of a job. 
The primary storage is used for moderate term end user data storage, and provides sufficient aggregate performance to support multiple users with various read/write operations. 
Unlike volatile scratch storage, primary cluster storage is backed up for disaster recovery to a separate and remote data storage array.
\subsubsection{Scratch Storage}
Colonial One's scratch storage is its computational work space; users load data to the cluster's scratch space in order to execute the desired computational analysis. 
Unlike traditional data storage, scratch storage on a cluster is accessed in parallel by multiple compute nodes, and must be high-throughput in nature (in gigabytes per second, 
rather than megabytes per second). The aggregate storage system utilization by end user jobs dictates the required aggregate performance of the scratch storage system. 
Additionally, HPC scratch storage utilizes a parallel file system, delivering parallel file access to compute nodes, and scaling performance throughput with the number of file 
system servers. There are two primary considerations for parallel files systems; Lustre~\cite{lustre}, an open source platform, and GPFS~\cite{gpfs}, 
a commercial platform owned and developed by IBM. 
In some cases in which there are many read/write operations acting on many small files, as is often the case in the life sciences, performance of the Lustre file system is less 
than optimal due 
to the overhead associated with look-ups made to the metadata server~\cite{lustreperf} 
while GPFS, a block-based file system, is not similarly impacted. Given the cost overhead of GPFS and the desire to 
balance performance over all probable use cases, Lustre was provisioned as scratch storage for Colonial One. 
Because the scratch storage is used for running jobs and only stores data "in play", it is not be backed up.
\subsubsection{Archival Storage}
Colonial One leverages Dell's Compellent storage array for archival storage and provides a long-term storage solution for data at rest. 
Archive storage is not generally accessible to the user community though shares may be purchased by researchers. 

\section{Project Budget}
The total initial project budget put forward had specified a 96-node FDR InfiniBand cluster design that included 32 nodes with dual NVIDIA 
Kepler K20 GPU cards with the additional rack space for expansion as other Schools and Institutes made investments and were on-boarded.

It is important to note that the server hardware, storage arrays, and networking 
equipment represent approximately 75\% of the initial investment. The remaining 25\% of the project budget was expended on software, implementation services, and installation 
materials. The budget did not include operating budget line items for staff and operating expenses in the initial budget.

\subsection{Resource Manager and Job Scheduler Software}
The cluster's resource manager and job scheduler work in concert to maximize the availability of the right hardware resources for the right end user. 
The resource manager monitors individual compute nodes for hardware configuration, health, and availability, in order to ensure the job scheduler only assigns jobs to nodes 
that are capable of processing an end user's job. The job scheduler maintains availablity of resources for certain users or groups, ensuring policies for priority or job 
preemption are accurately implemented and enforced.

Resource management and scheduling on Colonial One are handled by SLURM (Simple Linux Utility for Resource Management). Two resource managing and scheduling options were evaluated, 
LSF (IBM, proprietary) and SLURM  
and both are well-suited for Colonial One, \textit{e.g.}, both support fair-share scheduling, preemption, and backfilling. 
There were three factors driving our 
adoption of SLURM over LSF (or OpenLava, the open source alternative of LSF):
\begin{enumerate}
\item open source and lower maintenance costs  
\item large active user community
\item previously existing in-house familiarity and expertise 
\end{enumerate}

Colonial One features several SLURM partitions organized by hardware profile and time limits. There are long, 14-day CPU partitions and short, 2-day CPU partitions, and two 
7-day GPU partitions that allow users the option to run with or without Error-correcting code memory. There is also a big memory (2TB) partition as well as a 4-hour debug partition. 
\section{Investment and Expansion}
Investment and expansion is a partnership between the University, Columbian College of Arts and Sciences (CCAS)~\cite{ccas}, 
the School of Engineering and Applied Science (SEAS)~\cite{seas}, the
Computational Biology Institute (CBI)~\cite{cbi}, the Milken Institute School of Public Health (SPH)~\cite{milken} and individual faculty. 
Given the operational and long-term organizational efficiencies gained by centralizing HPC resources, the University would assume certain 
infrastructure costs to subsidize the initial deployment and ongoing operational expenses. Over time, as the initial capacity saturated,
data center hosting space and the backend infrastructure (interconnect network, storage, and enclosures) have been increased to support continued expansion.

\subsection{University Support}

\subsubsection{Data Center Hosting}
The University currently provides data center hosting and network connectivity on a per-rack and per-connection basis. Data center hosting and upstream network connectivity for 
the cluster is provided by the Division of IT (DIT)~\cite{dit} in the Ashburn, Virgina Enterprise Hall data center. CCAS worked with DIT's data center team to 
plan the power distribution and physical floor space to accommodate seven standard equipment racks. Both OVPR and DIT have requested 
that this chargeback be 
funded from University-level Facilities and Administrative indirect funds.

\subsubsection{Staffing}
New staff resources was one of the most critical success factors as well as the most difficult to fund long-term. In order to adequately staff an initiative of this size and scope 
several key positions are recommended, including one senior and one junior HPC systems administrator as well as one full-time equivalent HPC specialist from each of the of the constituent
Schools and Institutes to provide training and support to users and to escalate issues to admins when necessary. An email support system is implemented to centralize reporting and tracking 
of user issues. The services of an HPC management company, R Systems NA Inc.\cite{rsystems}, have been retained to facilitate issues resolution during periods when Colonial One is short-staffed.  

\subsubsection{Faculty/Research Group Support}
Many Universities have developed faculty contribution programs in order to provide a mechanism for individual faculty and research groups to deploy dedicated cluster nodes. 
These contribution programs allow for realization of economies of scale from pooled resources and streamlined selection, integration, and deployment of equipment included on 
research proposals. Additionally, these programs effectively increase researcher efficiency by leveraging and expanding pre-existing infrastructure, transferring the responsibility of 
maintaining and supporting HPC resources to dedicated IT staff.

Non-priority access to Colonial One is open to the University community at no charge to researchers. 
Faculty opting into the shared cluster model purchase individual compute nodes and receive in return priority scheduling proportional to their investment. Additionally, long-term archival 
storage is available for purchase. When not in use, prioritized resources are made available to the general user population.

\section{Schedule and Milestones}
The following is list of milestones in the development of a shared HPCC at GWU:
\begin{itemize}
\item 2013-Q2 Burn-in and first production operation 
\item 2015-Q2 Expansion from 96 nodes to the present 213-node configuration. 
\item 2016-Q2 Transition from Terascala Lustre to Intel Enterprise Edition Lustre
\item 2018-Q1 End of lifecycle, commence burn-in and production operation in next hardware phase 
\end{itemize}

\section{User Support and Engagement}
Building a successful HPCC requires building a knowledgeable user base which can be somewhat of a challenge for a fledgling HPCC. As noted, the Colonial One user base is diverse. See Table~\ref{tab:popdisp} for a breakdown of users by discipline and Appendix~\ref{jobdiversity}
for example use cases.  
\begin{table}
\centering
\caption{User Population by Discipline}
\begin{tabular}{|l|c|} \hline
Stakeholder& User Population (\%)\\ \hline
Arts \& Sciences & 32\\ \hline
Engineering \& Applied Sciences & 28\\ \hline
Medical, Business, Genomics& 26\\ \hline
Computional Biology& 8\\ \hline
Support Staff& 6\\ \hline
\hline\end{tabular}
\label{tab:popdisp}
\end{table}
Users come to HPC from a variety of backgrounds and
each brings a distinct set of expectations. There are several useful instruments that facilitate user education. One-on-one or group training is available to researchers by consulting with 
the HPC specialist from their respective School or Institute. A successful HPCC environment requires HPC specialists who are familiar with the research domains represented to facilitate 
onboarding as well as installation and support of domain-specific applications. These resources and staff positions are committed and funded by SEAS, CCAS, and DIT.

The Extreme Science and Engineering Discovery Environment (XSEDE)~\cite{xsede}, funded by the National Science Foundation, 
offers free HPC training to users, both self-paced online as well as monthly live streaming classes in partnership with Pittsburgh Supercomputing Center. XSEDE also sponsors the 
XSEDE Campus Champion program which positions representatives on campuses that can facilitate access to XSEDE resources and provide start-up allocations at some of the largest HPC centers 
including the Texas Advanced Computing Center, National Center for Supercomputing Applications, Pittsburgh Supercomputing Center, and the San Diego Supercomputer Center. 
Additionally, a website for 
Colonial One (https://colonialone.gwu.edu) provides useful information and examples as well as links to video tutorials. 
In the first two years of operation, monthly town hall-style user lunch meetings were hosted in which training was offered at various levels on a variety of topics and similar 
group training 
sessions are still available upon request. Other outreach opportunities for Colonial One include representation at University sponsored research showcases as well as data sciences and 
related user group events. 

Several metrics are tracked to ascertain the benefits users derive from a hosted shared HPCC at GWU. Users are asked to self-report any publications or grant awards they receive that 
leverage Colonial One. A data warehouse exists that supports drilling down into job histories, user training, data transfer and priority allocations which allows support staff to 
generate reports, usage characteristics by user, group, and department and target cases where additional training may be useful.  

\section{Conclusions}
%\end{document}  % This is where a 'short' article might terminate
Creating and maintaining a thriving HPC environment is essential for any research university and requires a partnership between administration, faculty, and staff. 
We describe a successful investment and cost-sharing model to support a variety of research groups with diverse needs. We investigated storage, networking, and computational 
capability appropriate for engineering, life sciences, and physical sciences as well as describe the selection criteria for resource management and scheduling. The long-term success also 
requires provisioning the necessary staff lines and we provide a model for domain-specific user support and engagement. Lastly, metrics for success include tracking user statistics and 
helpdesk reports, evaluating effectiveness of training, and grants and awards statistics for the research being enabled.

%ACKNOWLEDGMENTS are optional
\section{Acknowledgments}
We gratefully acknowledge Sean Connolly and Warren Santner who did much of the work to pioneer the implementation of a HPCC at The George Washington University.

\bibliographystyle{abbrv}
\bibliography{c1paper}  % sigproc.bib is the name of the Bibliography in this case
% ACM needs 'a single self-contained file'!
%\balancecolumns
\appendix
%Appendix A
\section{Applications and Use Cases}\label{jobdiversity}
As a shared-resource, Colonial One is required to perform well in a diverse set of applications. The following is a non-exhaustive list of some of the applications being supported by 
Colonial One. 
\begin{itemize}
\item Study structure of subatomic particles
\item Large-scale molecular dynamics simulations
\item Network analysis 
\item Drug design for cancer therapy 
\item Protein engineering for immune response against bacteria and viruses including HIV/AIDS
\item fMRI analyses of injured brains
\item Genome sequencing
\item Phylogenetic mapping of evolutionary traits
\item Satellite imagery
\item Population and census dynamics
\end{itemize}
%\balancecolumns % GM June 2007
% That's all folks!
\end{document}